# RFID-Cloud Integration for Smart Management of Public Car Parking Spaces

Umar Yahya
Motion Analysis Research Laboratory,
Islamic University in Uganda
Kampala, Uganda
umar.yahya@ieee.org

Ndawula Noah
Motion Analysis Research Laboratory,
Islamic University in Uganda
Kampala, Uganda
ndawulanoah15@gmail.com

Asingwire Hanifah
Motion Analysis Research Laboratory,
Islamic University in Uganda
Kampala, Uganda
hanifahmntz@gmail.com

Lubega Faham
Motion Analysis Research Laboratory,
Islamic University in Uganda
Kampala, Uganda
fahamumar@gmail.com

Abdal Kasule
Department of Computer Science and
IT, Islamic University in Uganda
Kampala, Uganda
abdal78@gmail.com

Hamisi Ramadhan Mubarak
Department of Technical and
Vocational Education, Islamic
University of Technology, Bangladesh
hamisiramadhan@iut-dhaka.edu

*Abstract* –Effective management of public shared spaces such as car parking space, is one challenging transformational aspect for many cities, especially in the developing World. By leveraging sensing technologies, cloud computing, and Artificial Intelligence, Cities are increasingly being managed smartly. Smart Cities not only bring convenience to City dwellers, but also improve their quality of life as advocated for by United Nations in the 2030 Sustainable Development Goal on Sustainable Cities and Communities. Through integration of Internet of Things and Cloud Computing, this paper presents a successful proof-of-concept implementation of a framework for managing public car parking spaces. Reservation of parking slots is done through a cloud-hosted application, while access to and out of the parking slot is enabled through Radio Frequency Identification (RFID) technology which in real-time, accordingly triggers update of the parking slot availability in the cloud-hosted database. This framework could bring considerable convenience to City dwellers since motorists only have to drive to a parking space when sure of a vacant parking slot, an important stride towards realization of sustainable smart cities and communities.

## I. INTRODUCTION

As urban areas continue to experience positive population growth and increased human activity, there is growing need for cities to effectively and efficiently manage public shared spaces[1]. The need for efficient management of shared spaces in urban areas is clearly advocated for by United Nations in her 2030 sustainable development goal 11 (SDG11) on sustainable smart cities and communities [2]. Particularly, management of public car parking spaces in urban areas, has received a lot of interest in recent years [3], [4], [13], [5]–[12].

Failure to effectively manage public parking spaces has been attributed to increased fuel consumption and thus increased carbon emissions as motorists drive through urban spaces in search for parking slots [8], [14], [15]. Not only does this contribute to the poor air quality in urban spaces, but also results in time wastage, factors that directly impact the economies of the affected urban communities [15], [16]. For instance, it has been reported that the average search time for a parking slot was about 20 minutes for motorists in several European urban areas[8]. Even as cities build more parking spaces, the lack of efficient management of the same renders their effort less useful in as far as management of traffic congestion in urban spaces is concerned. For example, it has also been reported that in 2010 about 30% of urban traffic in the USA was caused by motorists searching for parking space, resulting in about 4.8 billion hours of time wasted to traffic[8].

Therefore, part of the solution to problems arising from inefficient management of public car parking spaces lies in leveraging the various emerging technologies in the design and adoption of smart management frameworks [3], [17]–[19]. Previous related works have explored use of Internet of Things (IoT) frameworks [10]–[12], Artificial Intelligence[5], [7], [20]–[23], Wireless Sensor Networks [3], [24]–[29], and cloud computing[3], [26], [30], [31], among others. However, affordability and sustainability of frameworks presented in literature, remains a huge challenge to many Cities in developing countries, and hence the rationale for this current study. Thus the main objective of this work was therefore to design and build a low-cost framework for effective management of public car parking spaces that is affordable by Cities and urban areas in developing countries. Through the proposed framework, motorists are able to reserve parking space via a mobile or web-based application, while access to and out of the parking space is enabled using Radio Frequency Identification (RFID). Status of parking space availability is accordingly updated in the cloud database accessible to all users through the mobile/web application, in real-time.

## II. METHODS

The proposed low-cost framework consists of three main layers as illustrated in Fig. 1 below. These layers include, IoT Layer, Cloud Layer and Application Layer.

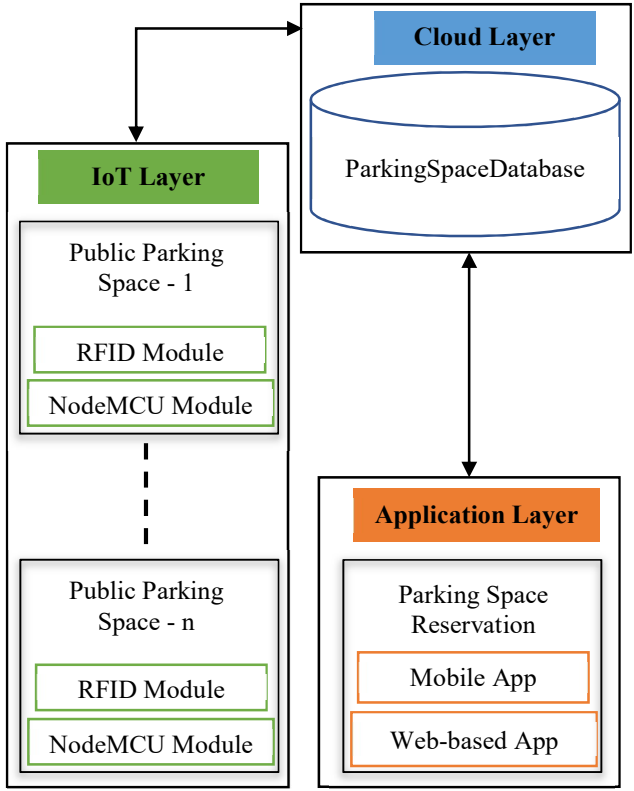

Fig. 1. Conceptual diagram of the proposed framework

### A. IoT Layer

As indicated in Fig.1, this layer resides in each registered public parking space (i.e public parking spaces 1, 2, ..., n) and constitutes an integrated sensing module (ISM). The ISM used in building the successful prototype presented in this work includes: -

RFID Module – The passive RFID RC522 module was used. It consist of a key chain, RFID card as well as the RFID reader which serves as the transceiver for reading the card [32]. Through an Arduino Microcontroller, authentication output of this module serves as input for the servo motor responsible for opening and closing the entry gate to the parking space.

NodeMCU ESP8266 Module – Using this integrated WiFi chip, the RFID authentication output is used to update the parking slot status in the cloud database accordingly, in real-time. Checking the cloud database to see whether the registered card holder had actually reserved a parking slot by the RFID interrogator component, and accordingly communicating any slot status changes, is all made possible through the NodeMCU module [33].

### B. Cloud Layer

As indicated in Fig. 1, this layer has to reside on the cloud in order for it to be accessible by motorists from any geographical location with Internet connectivity. It comprises of a relational car parking space database schema created in MySQL as illustrated in Fig 2 below. The database is updated with both motorist actions via mobile/web application at the application layer as well as at the IoT layer as per data transfer communications via NodeMCU module. The schema consists of three main entities: -

Parking Space Entity – This enables storage of details of each publicly available car parking space. Details include parking space ID, name, GPS coordinates for parking space location, possible maximum slots, number of slots currently reserved, number of slots currently occupied, a derived field for number of slots currently available, parking space administrator and their contact details, and whether parking is free of charge or otherwise. This implies that car parking spaces would have to be registered in this database first, before they can be accessed by interested motorists.

Motorists Entity – This enables storage of motorists' details. These include their full name, nationality, passport or national ID number, and contact details. All motorists intending to make use of the publicly available parking spaces in a city or urban space would therefore have to be registered first in this database.

Reservations Entity – This enables storage of reservation details such as RFID unique ID for the motorist, their national or passport ID, parking space ID, and time of slot reservation. The entity is populated and updated as per reservation actions performed by motorists through the application layer as well as by the RFID authentication output as transmitted through NodeMCU module.

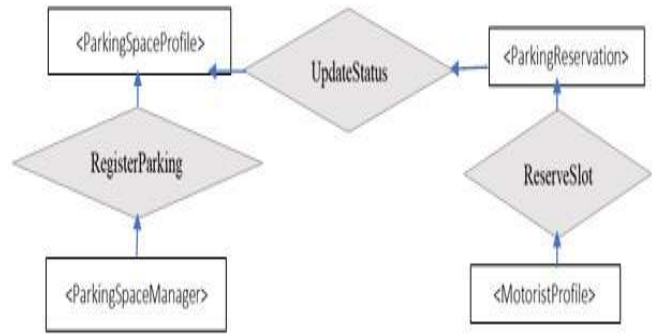

Figure 2. A partial representation of the entities in the cloud-hosted database

### C. Application Layer

This layer enables motorists to check for availability of parking space and complete reservations remotely, if needed. This was enabled both through a web-based application developed mainly using PHP and MySQL, as well as a mobile application for android smart device users. The applications have access to the car parking space database at cloud layer, which enables new motorists to first register before being able to complete any space reservations. Application users are able to see the status of each registered car parking space available to the public. Moreover, the real-time interaction between the IoT layer and the Cloud Layer, ensures that motorists are able to see the state of the parking spaces as is in real-time as illustrated in Fig. 1.

## III. IMPLEMENTATION RESULTS

This framework proposed in this work was successfully prototyped as explained in the Methods section of this paper. The following subsections demonstrate the successful prototype of the framework.

### A. IoT Layer Implementation

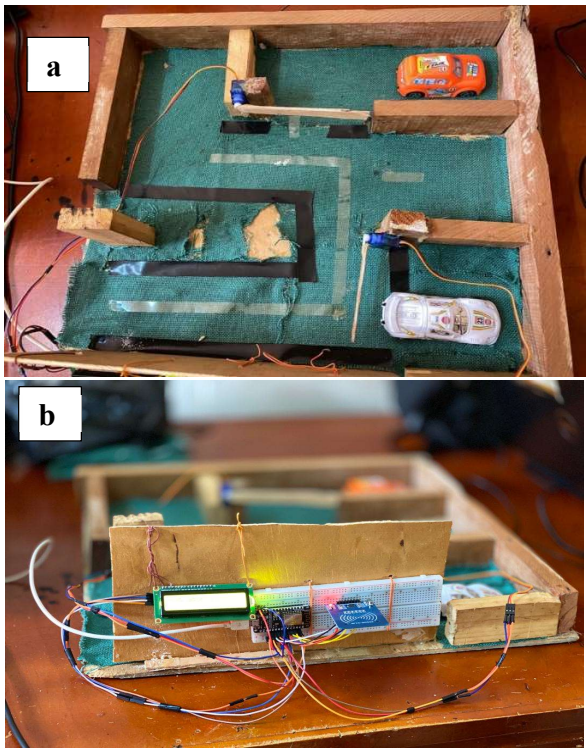

Fig. 3. A prototype of a public car parking space embedded with IoT layer

As demonstrated in Fig. 3 (a) above, a public parking space with two car slots was prototyped. Entry of cars into and exit from the parking slot is enabled the registered RFID cards of registered motorists are authenticated by the RFID reader seen in Fig. 3 (b). The RFID authentication output is used as input to allow the servo motor open or close the parking slot gate as seen in Fig. 3 (a) above. Acceptance or rejection following RFID authentication is also displayed to the motorist on an LCD screen as seen in Fig. 4 below.

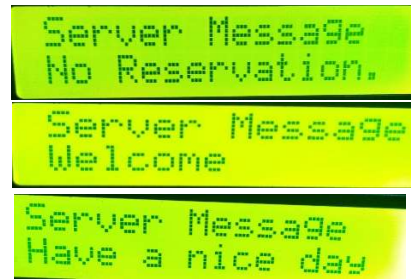

Fig. 4. RFID authentication feedback to motorist at parking lot

### B. Parking Space Reservation at the Application Layer

Using the web-based interface of the application layer, a registered motorist user is able to see the status of the parking lot as per Fig. 5 (a), and thereafter is able to reserve a slot by selecting from any of the vacant slot numbers. Reserving a slot would change its status in the database and recolor it orange as in Fig. 5 (b), while going on to occupy the reserved slot would also update its status in the database as well as recolor it red as seen in Fig. 5 (c) below.

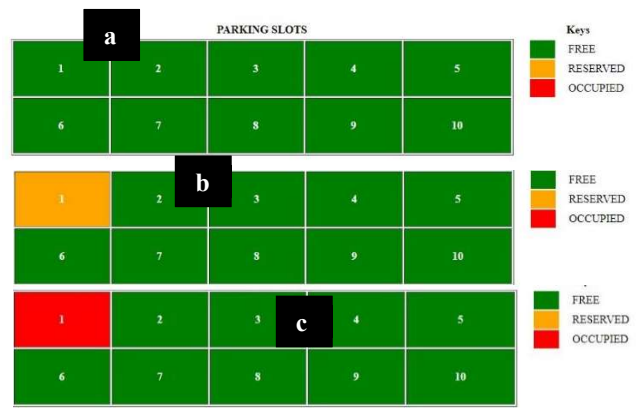

Fig. 5. Parking space reservation using web-based interface

Similar to web-based reservation explained above, reservation through the mobile application was also successfully implemented. The status of the reserved and/or occupied parking slot is updated in the database in the same manner illustrated in Fig. 5 above.

## IV. DISCUSSION AND FUTURE RECOMMENDATIONS

The successful implementation of the framework proposed in the current study adds to the growing list of smart management frameworks of Cities and urban areas as reported in several literature reviews [3], [4], [6], [34]–[36]. The use of low-cost technologies at the IoT Layer demonstrates the feasibility of proposed framework, particularly for application in Cities and urban areas of low income and developing countries as advocated for in [36]. Some of the previously proposed efficient frameworks such as those utilizing image processing [7] and deep learning [23] could be quite costly for low income Cities to afford, and hence hinder implementation.

By enabling registration of any publicly available parking space in this presented framework, authorities responsible for parking spaces in Cities and urban areas would experience significant improvements in management of these spaces as suggested in [19]. Moreover, other aspects such as revenue collection from these public spaces would also be made more efficient, considerably. City and urban area dwellers would also experience considerable reductions in accidents, delays, and carbon emissions as has been attributed to motorists movements in search for parking slots [8], [34], [37].

To implement the proposed framework in a real life setting in a City or urban space, network gateways would have to be installed at the IoT Layer to enable real-time transmission of aggregated NodeMCU data to the Cloud Layer. This is necessary because NodeMCU module alone would not be sufficient to transmit data over long distances so as to enable real-time interaction with the Cloud Layer [33]. This however would not be a serious limitation as several Cities even those in developing countries, are working towards installation of public Internet hotspots [38].

As was previously applied in a database-driven Neural Computing system [39], future improvements of the proposed framework could incorporate intelligent data analytics techniques at the Cloud Layer in order to continuously derive new insights in the huge volume of reservation data to be collected over time. Furthermore, the timestamps [40] of entry and exit moments of cars at parking lots could be used to facilitate express electronic payment, in cases where the parking slot is not available for free or where a motorist exceeds the specified free parking time for a given public parking space.

## V. CONCLUSION

This work has presented a successful implementation of a new proposed low-cost management framework for public car parking spaces in Cities and urban areas. Moreover, the framework presented in this work demonstrates its feasibility for deployment in low income Cities as well as most developing countries. This presented solution could go a long to enhancing efficiency in management of public shared spaces, a key indicator towards the realization of smartly managed cities as advocated for by the United Nations in its 2030 sustainable development goal on sustainable smart cities and communities. The framework further highlights the need for Africa, particularly, to continue leveraging ICTs in order to foster multi-sectoral development on the continent.